\documentclass[manuscript]{aastex}

\usepackage{color}

\shorttitle{Herschel/HIFI observations of hydrogen fluoride}
\shortauthors{Monje et al.}

\begin{document}


\title{Herschel/HIFI observations of hydrogen fluoride \\
     toward Sagittarius B2(M)}


\author{R. R. Monje, M. Emprechtinger, T. G. Phillips, D. C. Lis} 
\affil{California Institute of Technology, 1200 E. California Blvd., Pasadena, CA  91125-4700, USA}

\email{raquel@caltech.edu}
\author{P. F. Goldsmith}
\affil{Jet Propulsion Laboratory, California Institute of Technology, 4800 Oak Grove Drive, Pasadena CA 91109-8099, USA}

\author{T. A. Bell}
\affil{Departamento de Astrof\'{\i}sica, Centro de Astrobiolog\'{\i}a (CAB), INTA-CSIC, Crta. Torrej\'on km 4, 28850, Torrej\'on de Ardoz, Madrid, Spain}
\author{E. A. Bergin}
\affil{Department of Astronomy, University of Michigan, 500 Church Street, Ann Arbor, MI 48109,	USA}
\author{D. A. Neufeld}
\affil{Department of Physics and Astronomy, Johns Hopkins University, 3400 North Charles Street, Baltimore, MD 21218, USA}

\and

\author{P. Sonnentrucker}
\affil{Space Telescope Science Institute, 3700 San Martin Dr, Baltimore, MD 21218, USA}




\begin{abstract}

Herschel/HIFI observations have revealed the presence of widespread absorption by hydrogen fluoride (HF) $J=1-0$ rotational transition, toward a number of Galactic sources \citep{Son10,Neu10,Phi10}. We present observations of HF $J=1-0$ toward the high--mass star--forming region Sagittarius B2(M). 
The spectrum obtained shows a complex pattern of absorption, with numerous features covering a wide range of LSR velocities (-130 to 100 km s$^{-1}$). An analysis of this absorption yields HF abundances relative to H$_2$ of $\sim$1.3 $\times$ 10$^{-8}$, in most velocity intervals. This result is in good agreement with estimates from chemical models, which predict that HF should be the main reservoir of gas-phase fluorine under a wide variety of interstellar conditions. Interestingly, we also find velocity intervals in which the HF spectrum shows strong absorption features that are not present, or are very weak, in spectra of other molecules, such as $^{13}$CO (1--0) and CS (2--1). HF absorption reveals components of diffuse clouds with small extinction that can be studied for the first time. Another interesting observation is that water is significantly more abundant than hydrogen fluoride over a wide range of velocities toward Sagittarius B2(M), in contrast to the remarkably constant H$_2$O/HF abundance ratio with average value close to unity measured toward other Galactic sources. 

\end{abstract}



\section{Introduction}

Hydrogen fluoride (HF) is expected to be the main reservoir of fluorine (F) in the interstellar medium because of its unique thermochemistry. Fluorine, predominantly neutral in the diffuse ISM, forms a diatomic hydride molecule with a dissociation energy of 5.87~eV, greater than that of H$_{2}$ (4.48~eV). In consequence, fluorine atoms react exothermically with H$_{2}$, the dominant constituent of molecular clouds, forming hydrogen fluoride. Once the H$_{2}$/atomic H ratio exceeds $\sim$ 1, HF becomes the dominant reservoir of fluorine and is destroyed only slowly by photodissociation, with an estimated photodissociation rate of 1.17 $\times$ 10$^{-10}$~s$^{-1}$ for the mean IS radiation field, and by reactions with He$^{+}$, H$^{+}_{3}$, and C$^{+}$. Since HF formation is dominated by reaction of F with H$_2$, it is expected that the total H$_2$ and HF column densities track each other closely, with an estimated HF/H$_2$ abundance ratio within an order of magnitude from 3.6 $\times$ 10$^{-8}$, when H$_2$ and HF are the dominant reservoirs of gas--phase H and F nuclei, as shown by \cite{Neu05}. 

 The first detection of interstellar hydrogen fluoride was reported by \cite{Neu97} using the Long Wavelength Spectrometer (LWS) of the Infrared Space Observatory (ISO). The HF $J=2-1$ rotational transition was observed in absorption toward the far-infrared continuum source Sagittarius B2, at a relatively low spectral resolution (R = 9600). However, the need for strong radiative pumping, or extremely high gas density, to populate the HF $J=1$ level severely limits the utility of HF $J=2-1$ absorption as a probe of interstellar hydrogen fluoride (Neufeld et al. 2005). The HIFI instrument, on board the Herschel Space Observatory, has enabled for the first time, observations of the \emph{fundamental} $J=1-0$ rotational transition of HF at 1.232 THz to be performed at high spectral resolution (R $>$ 10$^6$). The line has been detected in environments as diverse as Orion~KL, OMC-1 \citep{Phi10} and in diffuse clouds on the line-of-sight toward W49N, W51 \citep{Son10} and W31C \citep{Neu10}. This transition is generally observed in absorption, as expected, due to its very large A--coefficient, A$_{10}$ = 2.42 $\times$ 10$^{-2}$~s$^{-1}$. Only an extremely dense region, with a strong radiation field, could generate enough excitation to yield an HF feature with a positive frequency-integrated flux \citep{Neu97}. 
The HF $J=1-0$ transition promises to be an extremely sensitive probe of the diffuse molecular gas along the line-of-sight to background far-infrared continuum sources. 
To date, such material has been studied in the submm regime primarily by means of CO rotational emission lines. However, \cite{Neu05} show in their models that in diffuse clouds of small extinction, the  predicted HF abundance can actually exceed that of CO, even though the gas-phase fluorine abundance is 4 orders of magnitude smaller than that of carbon. HF is a more reliable tracer than CO because the HF/H$_{2}$ ratio is constant, whereas CO/H$_{2}$ drops in clouds of small \textit{A$_{v}$} \citep{Neu05}. The high A--coefficient of HF $J=1-0$ results in simple excitation, with essentially all HF molecules being in the ground rotational state under conditions characteristic of the diffuse or even dense ISM. The HF absorption thus traces the total molecular column density along the line-of-sight, including very cold regions that may not be detectable in CO emission or other commonly used tracers of molecular hydrogen \citep[see,][]{Ber04}. HF depletion onto grain mantles has been predicted to occur in high-density regions. The apparent abundance of HF toward Orion~KL \citep{Phi10} is much lower compared to that characteristic of diffuse clouds (this may well be due to incomplete coverage of the continuum source by the absorbing material; Orion is an unusual source in that it appears to be heated from the front and therefore does not show absorption, in general, at submillimeter wavelengths). If we accept that all gas-phase fluorine is in the form of HF, then HF observations can reveal the extent to which volatile species are bound up in icy grain mantles. 

In this Letter, we report the result of the Herschel/HIFI observations of HF $J=1-0$ toward Sagittarius B2 (Sgr B2) as part of the HEXOS (Herschel/HIFI observations of EXtraordinary Sources) Open Time Key Program. Sgr B2 is a very massive (several million solar masses) and extremely active region of high-mass star formation with an extraordinary rich chemistry, at a projected distance of 100 pc from the Galactic Center and at 8.5 kpc from the Sun {\citep{Gol90}. Its strategic location and its strong submillimeter continuum flux, make Sgr B2 one of the best sources towards which to carry out absorption studies. The Sgr B2(M) line-of-sight includes essentially the entire path to the center of our Galaxy, allowing us to study simultaneously the physical and chemical conditions of the source itself, Galactic Center clouds and the spiral arm clouds. 

\section{Observations}

The $J=1-0$ transition of HF, with rest frequency 1232.4762~GHz \citep{nol87} was observed as a part of the full spectral scan of HIFI band 5a toward Sgr B2(M) ($\alpha_{J2000}$ = 17$^{h}$47$^{m}$20.350$^{s}$ and $\delta_{J2000}$ = -28$^{\circ}$23$\arcmin$03.00$\arcsec$) carried out on 2010 September 16. The dual beam switch (DBS) observing mode was used with reference beams located 3$\arcmin$ on either side of the source position along an East-West axis. We used the HIFI Wide Band Spectrometer (WBS) providing a spectral resolution of 1.1 MHz, corresponding to a velocity resolution of 0.27~km~s$^{-1}$ at the frequency of the HF $J=1-0$ line, over a 4~GHz Intermediate Frequency (IF) bandwidth. 

The data have been reduced using the Herschel Interactive Processing Environment (HIPE) \citep{Ott10} with pipeline version 5.2. Deconvolution of the double side band (DSB) data into a single-sideband (SSB) spectrum has been performed using the standard HIFI deconvolution task (\textit{doDeconvolution}) within HIPE. The resulting Level 2 SSB spectra were exported into FITS format for subsequent data reduction and analysis using the IRAM GILDAS package. Beam observations, reported on November 17 of 2010, towards Mars at 1243 GHz give a main beam ($\eta_{mb}$) and an aperture efficiency ($\eta_{A}$) of 0.61 and 0.54, respectively for the Horizontal (H) polarization, and $\eta_{mb}$~=~0.65 and $\eta_{A}$~=~0.58 for the vertical (V) polarization. The full width at half maximum (FWHM) HIFI beam size at the HF $J=1-0$ frequency is $\sim$17\arcsec.

\section{Results}

Figure~\ref{fig1} shows the spectrum of the ground-state rotational transition of HF toward Sagittarius B2(M). The spectrum presents a complex absorption line structure that covers a wide range of LSR velocities (-130 to 90~km~s$^{-1}$) with  weak emission tentatively detected at velocities in the range of 90--100~km~s$^{-1}$.
In the upper panel of Figure \ref{fig2} we present, for comparison, the spectrum of the 1$_{11}$--0$_{00}$ 1113.3430~GHz line of para-water, obtained using HIFI \citep{Lis10}. The para-water absorption spectrum shows a larger optical depth than that of HF at almost all LSR velocities, in contrast to what has been seen toward other Galactic sources, e.g. W51, W49N and W31C, where the H$_2$O/HF abundance ratio is remarkably constant ($\approx$ 1) and the spectra have similar profiles \citep{Son10,Neu10}. In Figure \ref{fig2}, lower panel, we compare the HF spectrum with $^{13}$CO (1--0) and CS (2--1) absorption spectra toward Sgr B2(M) line-of-sight \citep[observed with the IRAM 30-meter telescope;][]{Lis01}. The $^{13}$CO (1--0) and CS (2--1) spectra show absorption components in close correspondence to those seen in the HF absorption spectrum. Most of these absorption features have been attributed to well-known molecular gas clouds along the line-of-sight towards Sgr B2 \citep{Whi79, Gre94}. We can distinguish three main velocity intervals. The first interval, with velocities from 35 to 100 km s$^{-1}$, is mainly dominated by gas associated with the envelope of Sgr B2, which yields broad absorption centered near \textit{v}$_{LSR}$ $\sim$ 64 km s$^{-1}$. The second interval, with velocities from -10 to 35 km s$^{-1}$, shows absorption due to gas within a few hundred parsecs of the Sun, as well as gas within spiral arms at a galactocentric radius $\sim$ 5--8 kpc. The third interval, with velocities from -130 to -10 km s$^{-1}$, shows absorption associated with gas located within 1 kpc of the Galactic Center (\textit{v}$_{LSR}$ -130 to -50 km s$^{-1}$) and in foreground gas at a galactocentric radius 3--5 kpc (-50 to -10 km s$^{-1}$). 

We calculate the hydrogen fluoride column densities in the velocities ranges where lines are not saturated. First, we derive optical depths of the HF lines ($\tau$ = -ln[1-T$_{L}$/T$_{C}$], where T$_{L}$/T$_{C}$ is the line-to-continuum ratio), assuming that the foreground absorption completely covers the continuum source and that all HF molecules are in the ground state. We set a conservative lower limit of $\sim$ ln(10) for the optical depth since the observations are not sensitive to larger opacities. The resulting optical depth is shown in Figure \ref{fig1} (lower panel). We derive the HF column densities for each LSR velocity range using equation 3 in Neufeld et al. (2010). Table 1 gives the HF column densities in several velocity intervals associated with different molecular clouds. The H$_2$ column densities in the foreground gas are calculated based on the $^{13}$CO absorption data from \cite{Lis01}, assuming a CO fractional abundance for diffuse molecular clouds of $\sim$3$\times$10$^{-5}$ \citep{Son07} and a [$^{12}$CO/$^{13}$CO] ratio of 60 in the local gas in the Sagittarius arm (-10 -- 35 km/s) and 30 for the remaining velocities \citep{Lan90}. In their studies of fluorine chemistry in the interstellar medium, \cite{Neu09} predicted an N(HF)/N(H$_2$) abundance ratio of 3.6 $\times$ 10$^{-8}$. Our resultant HF column densities relative to that of H$_2$ are $\sim$ 1.3 $\times$ 10$^{-8}$, in good agreement with the chemical models.
 We also calculate the N(H$_2$O)/N(HF) ratio using the water column densities derived in \cite{Lis10} in the velocity intervals of -73 to -52 km s$^{-1}$, -12 to -7 km s$^{-1}$ and 27 to 35 km s$^{-1}$. The mean H$_2$O/HF column density ratio in those intervals is 13.8 with a standard deviation of 0.58. The resulting H$_2$O/HF ratio is thus significantly higher than that observed toward other submillimeter sources - such as W31C, W49N and W51  - where the H$_2$O/HF ratio is close to unity and shows a remarkably small dispersion (Neufeld et al. 2010; Sonnentrucker et al. 2010).  Our analysis indicates that the HF abundance along the sight-line to Sgr B2(M) is similar to that inferred toward the other sources, and that the larger H$_2$O/HF ratio in Sgr B2(M) reflects an enhanced H$_2$O abundance along this sight-line.

\section{Discussion}

The observational results shown in Figure \ref{fig1}, reveal HF $J=1-0$ absorption over a broad range of LSR velocities along the line-of-sight toward Sgr B2. The derived abundances of HF with respect to H$_2$, show a mean of 1.3 $\times$ 10$^{-8}$ with a dispersion of 0.26 $\times$ 10$^{-8}$. Based on the average gas-phase interstellar abundance in diffuse atomic gas clouds N$_F$/N$_H$ = 1.8 $\times$ 10$^{-8}$ \citep{Sno07}, our observations suggest a small F depletion factor of $\sim$ 3 along the line-of-sight to Sgr B2(M). \cite{Neu10} and \cite{Son10} estimated similar depletion factors toward G10.6-0.4 and W49N and W51, respectively. These results corroborate the theoretical prediction that HF is the main reservoir of gas-phase fluorine in a wide variety of interstellar conditions where the HF and H$_2$ column densities track each other closely, with the N(HF)/2N(H$_2$) ratio equal to the gas--phase elemental abundance of F relative to H. The HF abundance obtained here, which applies to relatively diffuse gas along the sight-line to Sgr B2(M), is roughly two orders of magnitude larger than that inferred by Neufeld et al. (1997) from observations of the HF $J=2-1$ line. This difference is likely explained by the fact that the $J=2-1$ absorption line probes much denser gas, close to the Sgr B2 core, where the dust radiation field is sufficient to populate the $J=1$ state of HF; at high densities, the depletion of fluorine onto grain surfaces is likely far more efficient, as recently suggested by Phillips et al. (2010) in their discussion of HF observed towards Orion-KL.

The HF spectrum correlates well with the $^{13}$CO absorption spectra. However, a cursory comparison of the two spectra reveals at least two significant features in the HF spectrum, at velocities of -55 km s$^{-1}$ and -15 km s$^{-1}$, that are not present, or are very weak, in the $^{13}$CO data. To our knowledge, the molecular cloud at LSR velocity of $\sim$~-55 km s$^{-1}$ has not been seen in prior molecular absorption line studies. Since the HF/H$_{2}$ abundance ratio is independent on \textit{A$_v$} \citep{Neu05}, the low N($^{13}$CO)/N(HF) ratio for these velocity intervals (see Table 1) indicates that HF is tracing a low density region having low extinction (A$_v$~$\leq$~1) in which the CO abundance drops rapidly due to photodissociation \citep[][]{Lee96}. Assuming for those molecular clouds a N(HF)/N(H$_{2}$)~=~3.6~$\times$~10$^{-8}$ and using the standard gas to dust ratio, N(H$_2$)/A$_v$~=~9.4~$\times$~10$^{20}$ molecules cm$^{-2}$~mag$^{-1}$, to transform visual extinction into molecular hydrogen column density \citep{Fre82}, we estimate a visual extinction of the clouds of $\leq$~0.25 and $\leq$~0.65 for the -55 and -15 km~s$^{-1}$ molecular clouds, respectively. This result is in agreement with the prediction from the chemical model  of \cite{Neu05}, for which HF abundance in diffuse clouds of small extinction can exceed that of CO, even though the gas-phase fluorine abundance is 4 orders of magnitude smaller than that of carbon.

Another interesting result is that water is significantly more abundant than hydrogen fluoride toward Sgr B2 over a wide range of velocities, in contrast to the remarkably constant H$_2$O/HF abundance ratio of order unity toward other Galactic sources. The high water abundance was also observed by \cite{Neu00} with their observations of the 1$_{10}$--1$_{01}$ pure rotational transition of H$_{2}$O toward Sagittarius B2 using the \textit{Submillimeter Wave Astronomy Satellite}, with an average N(H$_{2}$O)/N(H$_2$) of 6 $\times$ 10$^{-7}$. At LSR velocities associated to gas within 1~kpc of the Galactic Center (velocity interval -130 to -10 km s$^{-1}$), a higher H$_2$O/HF line ratio has also been observed toward other sources close to the Galactic Center, e.g. Sgr A+50 km s$^{-1}$ (Sonnentrucker et al., in preparation).

\acknowledgments

We thank, the referee Dr. Andrew Walsh, for his helpful comments and suggestions. HIFI has been designed and built by a consortium of institutes and university departments from across Europe, Canada and the United States under the leadership of SRON Netherlands Institute for Space Research, Groningen, The Netherlands and with major contributions from Germany, France and the US. Consortium members are: Canada: CSA, U.Waterloo; France: CESR, LAB, LERMA, IRAM; Germany: KOSMA,MPIfR, MPS; Ireland, NUI Maynooth; Italy: ASI, IFSI-INAF, Osservatorio Astrofisico di Arcetri-INAF; Netherlands: SRON, TUD; Poland: CAMK, CBK; Spain: Observatorio Astronomico Nacional (IGN), Centro de Astrobiología (CSIC-INTA). Sweden: Chalmers University of Technology-MC2, RSS \& GARD; Onsala Space Observatory; Swedish National Space Board, Stockholm University - Stockholm Observatory; Switzerland: ETH Zurich, FHNW; USA: Caltech, JPL, NHSC. Support for this work was provided by NASA through an award issued by JPL/Caltech. This research has been supported in part by the NSF, award AST-0540882 to the CSO. A portion of this research was performed at the Jet Propulsion Laboratory, California Institute of Technology, under contract with the National Aeronautics and Space Administration.

{\it Facilities:} \facility{Herschel/HIFI}, \facility{IRAM 30m Telescope}.

\clearpage

\begin{figure}
\centering
\includegraphics[angle=-90,scale=.50]{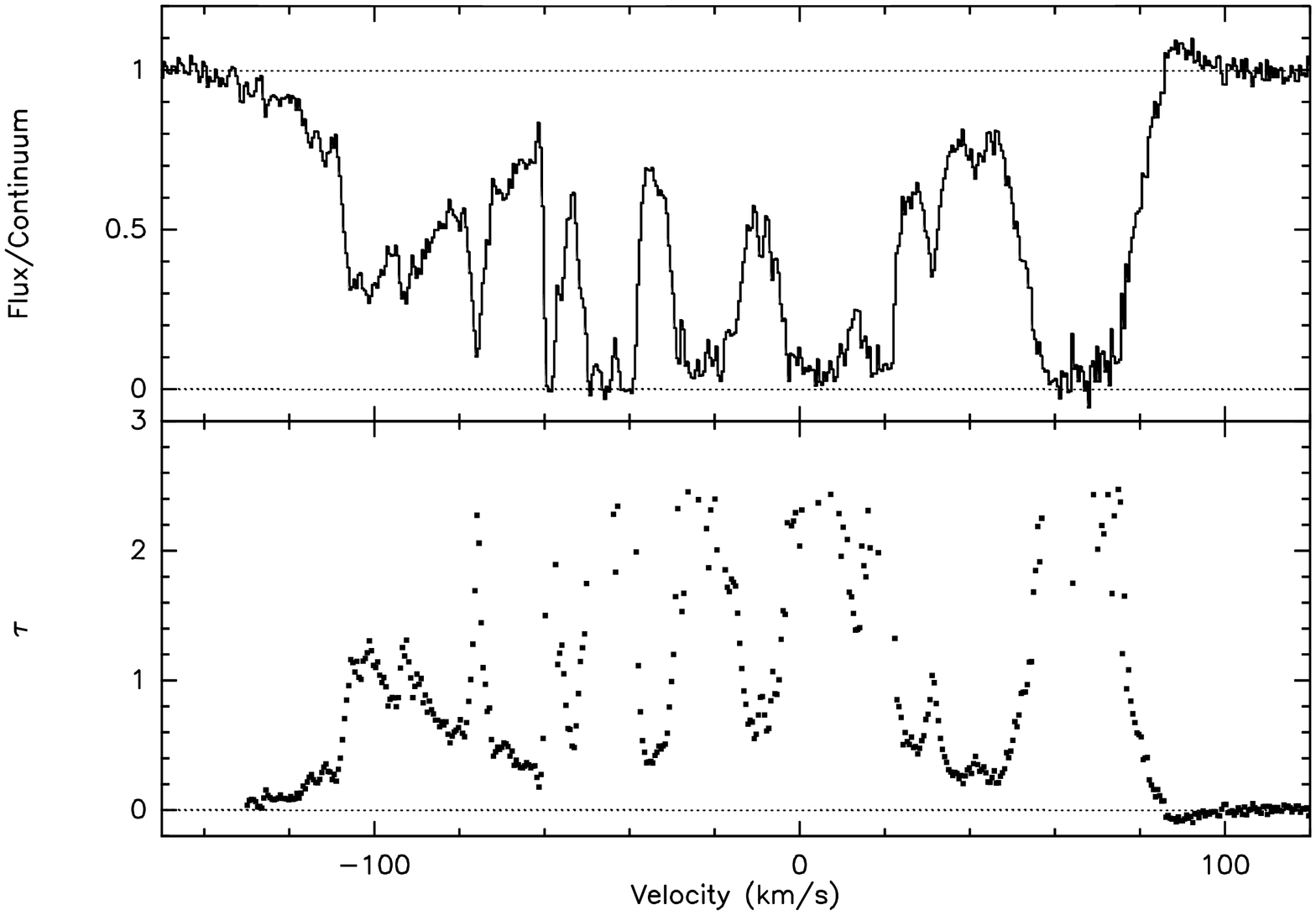}
\caption{(Upper) Spectrum of the ground-state transition of HF $J=1-0$ toward Sagittarius B2(M) normalized by the continuum. (Lower) Optical depth (for $\tau$ $\leq$ 2.5) as a function of LSR velocity. \label{fig1}}
\end{figure}

\clearpage

\begin{figure}
\centering
\includegraphics[angle=-90,scale=.60]{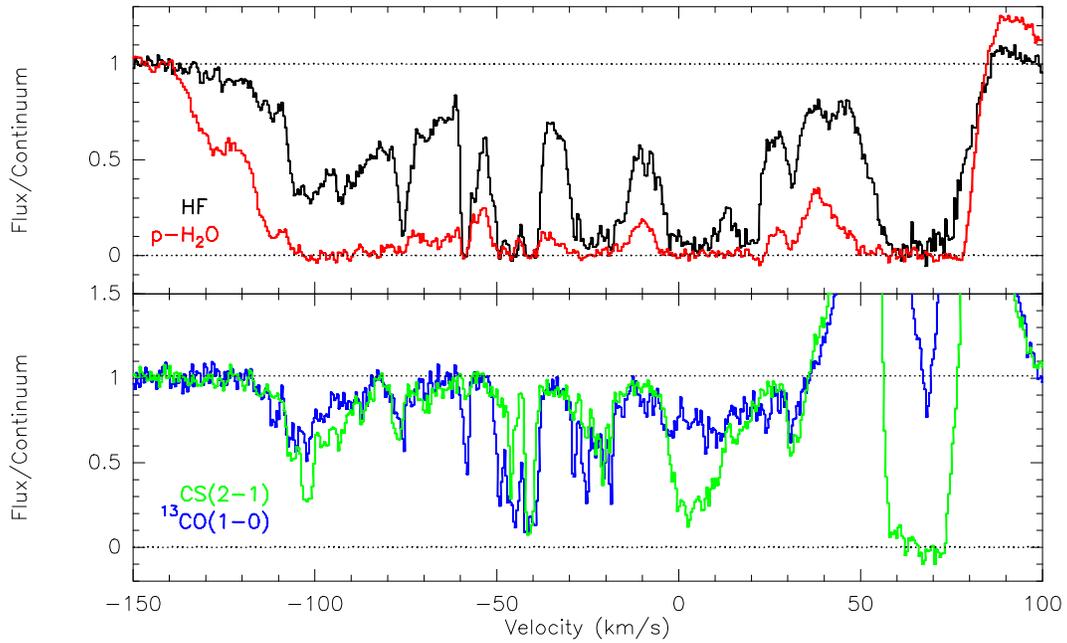}
\caption{(Upper) Spectra of the ground state of HF $J=1-0$ and p-H$_2$O $J=1_{11}-0_{00}$ lines (black and red histograms, respectively) toward Sagittarius B2(M) normalized by the continuum. (Lower) IRAM 30 m $^{13}$CO $J=1-0$ and CS $J=2-1$ spectra toward the same position (blue and green histograms, respectively).\label{fig2}}
\end{figure}

\clearpage
\begin{deluxetable}{lccccc}
\tabletypesize{\scriptsize}
\tablecaption{Column densities of HF and $^{13}$CO toward Sagittarius B2(M) and HF abundances.}
\tablewidth{340pt}
\tablehead{
\colhead{V$_{LSR}$}&N(HF)&N($^{13}$CO)&N(H$_2$)\tablenotemark{a}&HF/H$_2$&N($^{13}$CO)/N(HF)
\\
km s$^{-1}$ & cm$^{-2}$ & cm$^{-2}$& cm$^{-2}$&
}
\startdata
-115 to -60& 9.9~x~10$^{13}$ & 5.7~x~10$^{16}$&5.7~x~10$^{21}$& 1.7~x~10$^{-8}$&574\\[2pt]
-57 to -53& 8.3~x~10$^{12}$ & 2.7~x~10$^{14}$&$\cdots$&$\cdots$&33 \\[2pt]
-53 to -49 & 1.1~x~10$^{13}$ & 6.8~x~10$^{15}$&6.8~x~10$^{20}$ &1.5~x~10$^{-8}$ &650\\[2pt]
-35 to -28& 1.3~x~10$^{13}$  &9.3~x~10$^{15}$ & 9.3~x~10$^{20}$ & 1.4~x~10$^{-8}$& 706\\[2pt]
-18 to -11& 2.2~x~10$^{13}$  &5.3~x~10$^{15}$ & $\cdots$&$\cdots$& 236\\[2pt]
-10 to -5& 1.01~x~10$^{13}$  &4.8~x~10$^{15}$ & 9.6~x~10$^{20}$ & 1.1~x~10$^{-8}$&473 \\[2pt]
26 to 40& 1.7~x~10$^{13}$  &7.5~x~10$^{15}$ & 1.5~x~10$^{21}$ 	& 1.1~x~10$^{-8}$&445 \\[2pt]

\enddata
\tablenotetext{a}{The H$_2$ column densities are calculated based on the $^{13}$CO absorption data from \cite{Lis01}, assuming CO abundances of 3$\times$10$^{-5}$ \citep{Son07} and a [$^{12}$CO/$^{13}$CO] ratio of 60 in the local gas in the Sagittarius arm (-10 -- 35 km s$^{-1}$) and 30 in the remaining velocity intervals \citep{Lan90}. For those intervals with low N($^{13}$CO)/N(HF) ratio, the assumption of CO fractional abundances of 3$\times$10$^{-5}$ does not apply.}

\end{deluxetable}

\clearpage

\end{document}